\documentclass{article}

\PassOptionsToPackage{numbers, sort&compress}{natbib}

\usepackage[final]{neurips_2023}

\usepackage[utf8]{inputenc} 
\usepackage[T1]{fontenc}    
\usepackage{hyperref}       
\usepackage{url}            
\usepackage{booktabs}       
\usepackage{amsfonts}       
\usepackage{nicefrac}       
\usepackage{microtype}      
\usepackage{xcolor}         
\usepackage{amsmath}
\usepackage{amssymb}
\usepackage{graphicx}
\usepackage{enumerate}
\usepackage{enumitem}
\usepackage{bm}

\definecolor{hypercolor}{RGB}{174, 60, 60} 
\hypersetup{
  linkcolor  = hypercolor,
  citecolor  = hypercolor,
  urlcolor   = hypercolor,
  colorlinks = true
}

\title{Bayesian Simulation-based Inference for \\ Cosmological Initial Conditions}

\author{%
    Florian List  \\
  Department of Astrophysics \\
  University of Vienna, Austria\\
  \texttt{florian.list@univie.ac.at} \\
  \And
  Noemi Anau Montel \\
  GRAPPA Institute \\
  University of Amsterdam, The Netherlands \\
  \texttt{n.anaumontel@uva.nl} \\
  \And
  Christoph Weniger \\
  GRAPPA Institute \\
  University of Amsterdam, The Netherlands \\
  \texttt{c.weniger@uva.nl} \\
}

\begin{document}
\maketitle

\begin{abstract}
Reconstructing astrophysical and cosmological fields from observations is challenging. It requires accounting for non-linear transformations, mixing of spatial structure, and noise. In contrast, forward simulators that map fields to observations are readily available for many applications. 
We present a versatile Bayesian field reconstruction algorithm rooted in simulation-based inference and  enhanced by autoregressive modeling. The proposed technique is applicable to generic (non-differentiable) forward simulators and allows sampling from the posterior for the underlying field.
We show first promising results on a proof-of-concept application: the recovery of cosmological initial conditions from late-time density fields.

\end{abstract}

\section{Introduction} \label{sec:intro}

Recent developments in simulation-based machine learning are increasingly used for tackling difficult astrophysical and cosmological data analysis challenges~\cite[\textit{e.g.},][]{ Mishra-Sharma:2021oxe,
Cole:2021gwr, Alvey:2023npw, AnauMontel:2023stj, AnauMontel:2022ppb, Montel:2022fhv, Alsing:2017var, Alsing:2019dvb, Alsing:2019xrx, Makinen:2021nly, Modi:2023drt, Modi:2023llw, Barrue:2023ysk, Heinrich:2023bmt, Lin:2022ayr, Gagnon-Hartman:2023soa, Poh:2022ife, Lemos:2022kua, Brehmer:2019jyt, Alvey:2023naa, Bhardwaj:2023xph, Alvey:2023pkx, 
Karchev-sicret, Crisostomi_2023, Campeau-Poirier_2023, coogan2022walks, Hahn_2022simbig, Jeffrey_2020, Mishra_Sharma_2022, Saxena:2023tue}.
While simulation-based inference (SBI) has primarily been employed to solve relatively low-dimensional ($\lesssim 50$-dimensional) parameter estimation tasks~\cite{AnauMontel:2023stj, Alvey:2023naa, Alsing:2019xrx}, it has yet to cover 
higher-dimensionality problems like image reconstruction, which are an essential component in astrophysical and cosmological data analysis. 

Here, we focus on the recovery of cosmological initial conditions from late-time density fields. This task is a challenging test case for new algorithms thanks to the non-linear, non-local mapping from the Gaussian target to the observation.  Cosmic inflation predicts the density in the early Universe to be highly homogeneous, with tiny density fluctuations that are extremely well described as a Gaussian random field. These density perturbations then gradually grow over cosmic time due to gravity and eventually collapse into the non-Gaussian ``Cosmic Web'' structure observed today \cite{Bond1996}. 
The reconstruction of the initial density field from late-time observations is an ill-posed problem (the early-to-late mapping is not injective on small scales, ~\cite[\textit{e.g.},][]{Brenier2003}). Therefore, there is an entire {\it distribution} of possible initial conditions consistent with a given late-time density field. 

\paragraph{Our contribution.}  We frame the task of field reconstruction (or, viewed from a non-physical point of view, image reconstruction) as a parameter inference problem.  We combine the power of SBI in solving parametric inverse problems together with the scalability offered by autoregressive models. Autoregressive models have established their versatility in tackling high-dimensional distribution estimation tasks by breaking down the joint distribution into a product of conditionals \cite{Papamakarios_2018_masked, Uria_2016}, and have been successful in conditional image modelling \cite{van2016pixel}. Additionally, we employ a Gibbs sampling algorithm based on exact data augmentation (GEDA) \cite{marnissi2019geda} to efficiently sample image parameter posteriors.
We will formulate our method in a generic way to emphasize its applicability to a wide range of field/image reconstruction problems. Importantly, our approach accommodates arbitrary \textit{non-differentiable} forward simulators.

\paragraph{Related work.} The problem of inferring cosmological initial conditions has been studied since the late 1980s (see \textit{e.g.}\ Refs.~\cite{croft1997reconstruction, Frisch2002, Gramann1993, Nusser1992, peebles1989, weinberg1992} for classical papers), for instance by applying the least-action principle or using optimal transport. In the last decade, Bayesian models have been formulated for this task ~\cite[\textit{e.g.},][]{Jasche2013}, many of which rely on differentiable forward models \cite{Li2022, modi2021flowpm}, in conjunction with Hamiltonian Monte Carlo sampling. 
Recently, machine learning methods such as convolutional neural networks \cite{Shallue2023}, variational inference \cite{Modi2023}, recurrent inference machines \cite{modi2021cosmicrim}, and score-based modeling \cite{legin2023posterior} have also been explored in this context. 

\section{Methodology} \label{sec:method}

\paragraph{Background and problem setup.} SBI methods tackle statistical inverse problems by estimating posterior distributions from model-simulations. These methods do not require explicit modeling of the data likelihood, but instead access the information within the likelihood indirectly via a stochastic simulator, which maps input parameters to simulated data. Among various SBI algorithms (see Ref.~\cite{Cranmer_2020} for a review and Ref.~\cite{Lueckmann_2021} for benchmarks), we will focus on neural ratio estimation (NRE),\footnote{~We use \texttt{swyft} NRE implementation that can be found at \url{https://github.com/undark-lab/swyft}.
} which rephrases posterior estimation into a binary classification problem \cite{hermans2020likelihoodfree, Miller_2021}.

Let us assume we have a simple hierarchical simulator
\begin{equation}
    p(\bm x| \bm z)p(\bm z)
\end{equation}
where $\bm x\in \mathbb{R}^{N \times N}$ is observed and $\bm z\in \mathbb{R}^{N \times N}$ are image parameters (pixel values). Here, $p(\bm x|\bm z)$ can include non-linear, non-local transformations and non-Gaussian noise, and it is only implicitly defined through a forward simulator. We will discuss how we can, for a given observation $\bm x_o$, estimate an approximate but computationally efficient Gaussian likelihood that locally resembles $p(\bm x_o|\bm z)$ for a target observation $\bm x_o$. 
A fast surrogate can then be leveraged for downstream image analysis tasks.

\paragraph{Likelihood estimation.}
Firstly, in order to obtain locally optimal data summaries $\bm s(\bm x)$ for the image reconstruction task, we use NRE to estimate the marginal, pixel-wise ratios\footnote{~In this work we use the following notation for ratio estimators $\tilde r(a;b) = \frac{p(a,b)}{p(a)p(b)} = \frac{p(a|b)}{p(a)}$. If necessary, multiple variables are comma separated,
for example $\tilde  r(a; b, c)=\frac{p(a,b,c)}{p(a)p(b, c)}$.}
\begin{equation}
    \tilde r(s_i(\bm x); z_i) \equiv \frac{p(s_i , z_i)}{p(s_i)p(z_i)} \;.
\end{equation}
We assume here that both the joint and marginal distributions can be approximated as Gaussians, whereas the mapping $s_i(\bm x)$ is an arbitrary learnable function (usually a neural network). 
Means and covariances are estimated on-the-fly during training (similarly to batch normalization \cite{Ioffe2015}) and are not represented as learnable parameters.
Secondly, in order to obtain an estimate of the joint likelihood $p(\bm x|\bm z)$, we proceed as follows.  We split the problem auto-regressively along the observation axis, and we use NRE to estimate
\begin{equation} \label{eq:autoregressive}
\begin{split}
     \frac{p(\bm x| \bm z)}{p(\bm x)} &\simeq \frac{p(\bm s(\bm x)| \bm z)}{p(\bm s(\bm x))} = \prod_{i=1}^{N^2} \frac{p(s_i| \bm s_{1:i-1}, \bm z)}{p(s_i| \bm s_{1:i-1})} \\
     &\simeq \prod_{i=1}^{N^2}  \tilde r(s_i; l_i, z_i) = \prod_{i=1}^{N^2} \frac{p(s_i, l_i, z_i)}{p(s_i)p(l_i, z_i)} \;,
\end{split}
\end{equation}
where we have introduced $l_i = (\bm L(\bm s))_i$ with $\bm L$ a (generally non-linear) autoregressive function.
Again, we assume that the individual functions $p(s_i, l_i, z_i)$ are (three-dimensional multivariate) Gaussians.

By rewriting the above components (for a complete derivation and definitions of $\bm Q_{\text{like}}$ and $\bm b$ see Appendix~\ref{apx:info}), we can obtain the likelihood function in the so-called \emph{information form},
\begin{equation}\label{eq:likelihood}
    \ln p(\bm x|\bm z) = -\frac12 \bm z^T \bm Q_{\text{like}} \bm z + \bm b \bm z + C(\bm s) \;.
\end{equation}
In this paper, we assume that the precision matrix $\bm Q_{\text{like}}$ is diagonal. However, correlations between data summaries $\bm s$ are accounted for through the autoregressive function mentioned above; also, each component $s_i(\bm x)$ of the data summary may depend on all components of $\bm x$ and thus accounts for cross-pixel information.

To enhance the robustness of Bayesian approaches in data analysis, frequently likelihood tempering techniques are employed that result in conservative estimates \cite{holmes2017tempering, Jasche_2019}. 
Tempering the likelihood amounts to raising it to a fractional power $\gamma \in [0, 1]$, leading to progressively coarser posterior samples, with the extreme case of $\gamma=0$ ($\gamma = 1$) corresponding to samples drawn from the prior (posterior) distribution.

\paragraph{Posterior sampling.}
In our Bayesian framework and assuming Gaussian distributions, we combine the estimated likelihood with the known prior to derive the posterior in the information form
\begin{equation} \label{eq:post}
    \ln p(\bm z|\bm x) = -\frac12 \bm z^T \underbrace{(\bm Q_{\text{like}}+\bm Q_{\text{prior}})}_{\bm Q_{\text{post}}} \bm z + \bm b \bm z + C'(\bm s) \;, 
\end{equation}
where we have assumed a zero-mean prior, consistent with the physical problem we study in Sec.~\ref{sec:exp}.
We use the conjugate gradient (CG) algorithm\footnote{~We use a slightly modified implementation of the preconditioned CG algorithm from \url{https://github.com/sbarratt/torch_cg}.} 
to compute the maximum-a-posteriori (MAP) estimate $\bm z_{\text{MAP}}$ of the image parameters $\bm z$ by solving the linear system $\bm Q_{\text{post}} \bm z = \bm b$.
The surrogate Gaussian posterior distribution is then given by $p(\bm z | \bm x) = \mathcal{N}(\bm z_{\text{MAP}}, \bm Q_{\text{post}}^{-1})$.
Once we have the posterior distribution in the above form, we simply have to sample from it.
Various techniques have been presented over time to tackle the problem of efficient sampling from a high-dimensional Gaussian distribution (for a recent review, see Ref.~\cite{vono2020highdimensional}). 
To obtain the Gaussian posterior samples, we use a Gibbs sampler based on the generalized exact data augmentation algorithm 
(GEDA)~\cite{marnissi2019geda}. GEDA solves the problem of high-dimensional Gaussian sampling specifically for distributions whose precision matrix can be expressed as $\bm Q = \bm Q_1 + \bm Q_2$ by exploiting specific properties of $\bm Q_1$ and $\bm Q_2$ (for details see Appendix \ref{apx:geda}). Crucially, $\bm Q_{\text{post}}$ satisfies these constraints.

\section{Experiment} \label{sec:exp}
\begin{figure}
    \centering
    \resizebox{\textwidth}{!}{\includegraphics{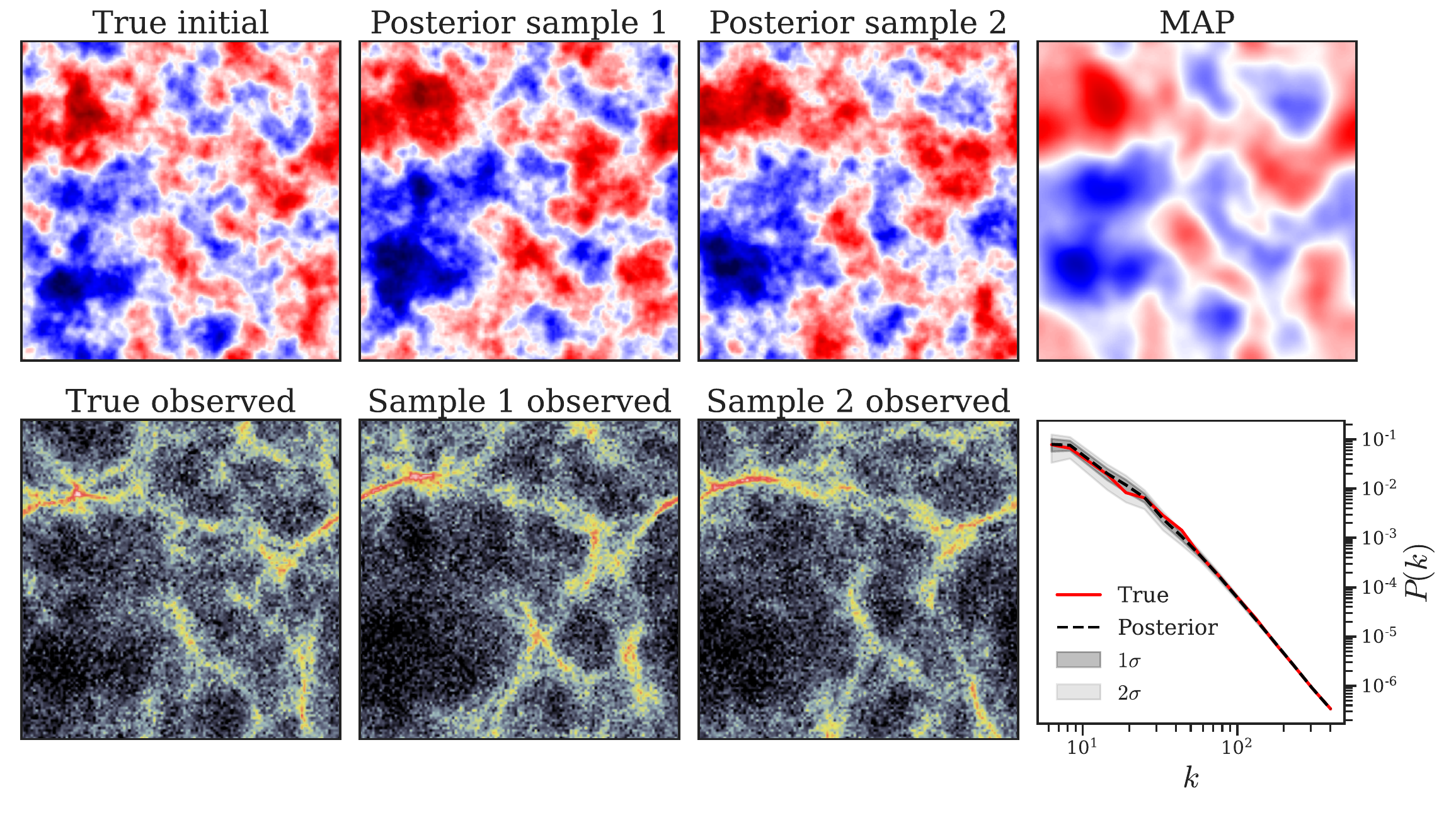}}
    \caption{Reconstruction of cosmological initial conditions in 2D. {\it Top left:} True initial density~${\bm z}_o$. {\it Bottom left}: True observation ${\bm x}_o$, i.e.\ the (logarithm of the) late-time density evolved from the initial conditions ${\bm z}_o$, corrupted by uncorrelated Gaussian noise. {\it Top center:} Two samples drawn from the posterior $p(\bm z | \bm {\bm x}_o)$. {\it Bottom center:} Observations computed from the posterior samples shown above. {\it Upper right:} Maximum-a-posteriori probability (MAP) estimate ${\bm z}_{\text{MAP}}$ of the initial conditions. {\it Bottom right:} Distribution of the reconstructed initial power spectrum.}
    \label{fig:results}
\end{figure}

To demonstrate the efficiency of our method, we apply it to the task of reconstructing the initial conditions of the Universe. In this proof-of-concept study, we consider the two-dimensional case and assume Einstein--de Sitter cosmology (i.e.\ non-relativistic, collisionless matter only). While our framework readily supports marginalizing over image parameters such as the power spectra of the target fields, we use a fixed power-law power spectrum for the initial density contrast $\bm z = \bm \delta_{\mathrm{ini}} \in \mathbb{R}^{128 \times 128}$, which is the target of our inference.
As a forward model, we use second-order Lagrangian perturbation theory (2LPT, see Refs.~\cite{Bouchet1992, Buchert1993} and Appendix~\ref{apx:2lpt}) and evolve the initial density to a time when non-linear structures have formed. The observation is then given by $\bm x = \log_{10}[1.1 + {\bm \delta}_{\mathrm{final}}] + \bm \varepsilon$, where ${\bm \delta}_{\mathrm{final}}$ is the density contrast at final time, the logarithm is applied element-wise, and we add $\bm \varepsilon \sim \mathcal{N}(\bm 0, \sigma^2 \bm I)$ with $\sigma = 0.15$ as a simplistic model for observational noise.

To obtain pixel-wise summaries $\bm{s}(\bm x)$, we use a standard U-Net \cite{ronneberger2015u}. In our current implementation, we use an autoregressive convolution ~\cite[\textit{e.g.},][]{van2016pixel} as the autoregressive function $\bm L$. In our experiments, we observed that overly small or large kernel sizes result in somewhat biased posteriors. Interestingly, we obtain the best results with a physically motivated kernel size for the convolution, i.e.\ when choosing it in such as way that the ``radius of influence'' of the autoregressive convolution (i.e.\ roughly half the kernel size) matches the typical distance that particles (or fluid elements)
have traveled by the end time -- a quantity known as displacement. Specifically, the mean displacement in our case is $\sim 5$ pixels, so we take the kernel size to be $11 \times 11$ pixels. While we find this choice to be optimal, slight bias still occurs occasionally, potentially due to the spatial heterogeneity of the displacements. Therefore, to be conservative, we use a tempered likelihood with $\gamma = 0.5$, trading some of the statistical constraining power of our method for increased unbiasedness. The importance of the specific choice of $\bm L$ (and in particular its radius of influence) indicates that a locally adaptive or multiscale approach for $\bm L$ is a promising avenue for future exploration. We train our model on 1000 $(\bm x, \bm z)$-pairs for 15 epochs, which only takes $\sim 3$ minutes on a laptop GPU for this experiment. Obtaining the MAP estimate with the CG algorithm takes less than 3 seconds. To obtain posterior samples with GEDA, we perform 300 sampling steps, which takes $<$ 1 second per sample (and multiple samples can be drawn in batches).

The left column of Fig.~\ref{fig:results} shows a target initial density contrast ${\bm z}_o$ ({\it top}), together with a resulting noisy observation at late time ${\bm x}_o$ ({\it bottom}). In the two center panels, we plot samples drawn from the (tempered) posterior $p(\bm z | {\bm x}_o)$, together with an observation of each sample. The upper right panel shows the MAP estimate ${\bm z}_{\text{MAP}}$. The initial density field is faithfully reconstructed on large scales whereas, as expected, small-scale information remains unconstrained. Finally, the lower right panel shows that the power spectra of the posterior samples are consistent with the target field. The excellent agreement between reconstructed and true power spectra on small scales is aided by the fact that the power spectrum (which is fixed in our example) directly enters the GEDA sampling (see Eq.~\ref{eq:precision_split} in Appendix \ref{apx:geda}, where the prior precision matrix ${\bm Q}_2$ is diagonalized by the Fourier transform, with the power spectrum on the diagonal of ${\bm D}_2$). We will present a detailed quantitative analysis of the results obtained with our framework in a separate publication.

\section{Discussion and Conclusion} \label{sec:conc}
We have introduced a framework for Bayesian field/image reconstruction by combining SBI, autoregressive modeling, and a Gibbs sampling algorithm based on exact data augmentation (GEDA). We presented promising results for a toy example related to reconstructing the initial conditions of the Universe. In view of its \textit{remarkable speed, low simulation costs, and the fact that it works for general non-differentiable simulators}, we expect our method to be capable of handling significantly higher-dimensional problems, including 3D cosmological simulations.

There are multiple promising avenues to be explored in future work. For many applications---such as the reconstruction of cosmological initial conditions---formulating the problem in Fourier space can be expected to produce significantly tighter posteriors, as the input-to-output mapping is exactly invertible on large to intermediate scales, and the stochasticity of the reconstruction only affects small scales. Alternatively, wavelets ~\cite[\textit{e.g.},][]{graps1995introduction} or other multiscale techniques could be exploited.
In this context (but also more generally), other choices for the autoregressive function $\bm L$ are worthwhile exploring, which could, for instance, implement the hierarchy between the different scales. 
Another promising enhancement involves harnessing the sequential aspect of SBI techniques. In principle, a viable strategy is to employ an adaptive scheduler to control the value of $\gamma$ to draw new training samples for sequential inference rounds.
Moreover, being SBI-based, our proposed method is capable of marginalizing over cosmological parameters or, alternatively, infer them in addition to the phases of the initial random field.

Let us also comment on the current limitations of our method: while the assumption of Gaussianity for the prior is justified for reconstructing cosmological initial conditions, using a Gaussian likelihood is an approximation whose justification depends on the specific task at hand. We take cross-pixel information into account through the summary statistics and an autoregressive function, but we currently model the precision matrix of the likelihood as being diagonal. In addition, the susceptibility of our framework to issues encountered in the context of autoregressive models such as exposure bias \cite{bengio2015scheduled} merits further investigation.

Finally, we remark that the forward model generating ``observations'' $\bm x$ does not necessarily need to be a physical one, and our framework also holds great promise for generic image reconstruction problems consisting in inferring one image from another.

\begin{ack}
This work is part of a project that has received funding from the European Research Council (ERC) under the European Union’s Horizon 2020 research and innovation program (Grant agreement No. 864035 -- UnDark). FL thanks Oliver Hahn and Cornelius Rampf for many insightful discussions.
\end{ack}

\bibliography{references}

\begin{appendix}

\section{Appendix: Derivation of the likelihood in the information form} \label{apx:info}

In this appendix we show how to derive the quadratic term $\bm Q_\text{like}$ and the linear term $\bm b$ of the likelihood in the information form (Eq.~\ref{eq:likelihood}) starting from Eq.~\ref{eq:autoregressive}. Having approximated $p(s_i, l_i, z_i)$ with a 3-dimensional Gaussian distribution, we can write the log-likelihood as
\begin{equation}
    \ln p(\bm s(\bm x)| \bm z) = \sum_{i=1}^{N^2} \ln \frac{p(s_i, l_i, z_i)}{p(l_i, z_i)} = \sum_{i=1}^{N^2} \ln \frac{\mathcal{N}(\mu_{(s_i, l_i, z_i)}, \Sigma_{(s_i, l_i, z_i)})}{\mathcal{N}(\mu_{(l_i, z_i)}, \Sigma_{(l_i, z_i)})} \;.
\end{equation}
Let us consider just one element $i$ and drop the index $i$ for simplicity of notation. Expanding the above expression for one component, we obtain
\begin{equation}\label{eq:matrix}
\begin{split}
    & -\frac12
    \begin{pmatrix}
    s-\mu_s \\
    l-\mu_l \\ 
    z-\mu_z \\
    \end{pmatrix}^T\hspace{-3mm}
    \underbrace{
    \begin{pmatrix}
    \Sigma_{ss} & \Sigma_{sl} & \Sigma_{sz} \\
    \Sigma_{ls} & \Sigma_{ll} & \Sigma_{lz} \\
    \Sigma_{zs} & \Sigma_{zl} & \Sigma_{zz}
    \end{pmatrix}^{-1}}_{
    \equiv Q' = 
    \begin{pmatrix}
    Q'_{ss} & Q'_{sl} & Q'_{sz} \\
    Q'_{ls} & Q'_{ll} & Q'_{lz} \\
    Q'_{zs} & Q'_{zl} & Q'_{zz}
    \end{pmatrix}
    }\hspace{-3mm}
    \begin{pmatrix}
    s-\mu_s \\
    l-\mu_l \\ 
    z-\mu_z 
    \end{pmatrix}
    +\frac12
    \begin{pmatrix}
    l-\mu_l \\ 
    z-\mu_z 
    \end{pmatrix}^T\hspace{-3mm}
    \underbrace{\begin{pmatrix}
    \Sigma_{ll} & \Sigma_{lz} \\
    \Sigma_{zl} & \Sigma_{zz}
    \end{pmatrix}^{-1}}_{
    \equiv Q'' = 
     \begin{pmatrix}
    Q''_{ll} & Q''_{lz} \\
    Q''_{zl} & Q''_{zz}
    \end{pmatrix}\;
    }\hspace{-3mm}
    \begin{pmatrix}
    l-\mu_l \\ 
    z-\mu_z
    \end{pmatrix} 
    \;.
\end{split}
\end{equation}

For a specific observation $\bm x$ we can compute $\bm s$ and therefore $\bm l$, hence individual components $s$ and $l$ (where we have dropped the index $i$). We can then read off the expression in Eq.~\ref{eq:matrix} the terms that depend on $z$. We group the rest into an irrelevant constant $C(s)$.
In this way, the \textit{quadratic term} in $z$ in Eq.~\ref{eq:matrix} is
\begin{flalign*}
     -\frac12 Q_{zz}z^2 &\quad\text{with}\quad Q_{zz} \equiv Q'_{zz} - Q''_{zz} \;.
\end{flalign*}
Each $Q_{zz}$ is a diagonal entry of $\bm Q_\text{like}$ in Eq.~\ref{eq:likelihood}, while non-diagonal components are 0 by construction. In principle, correlations between pixels in the data $\bm x$ can be modeled and we plan to explore it in future work.
The \textit{linear term} in $z$ in Eq.~\ref{eq:matrix} is
\begin{equation}
    b  \equiv \left[-(s-\mu_s) Q'_{sz} -(l-\mu_l) Q_{lz} + \mu_z Q_{zz}\right] \;.
\end{equation}
Each component $b$ composes the likelihood linear term $\bm b$.

\section{Appendix: Generalized Exact Data Augmentation {(GEDA)}}\label{apx:geda}

In general, data augmentation approaches target precision matrices of the form $\bm Q = \bm Q_1 + \bm Q_2$, which naturally arise from the statistical model under investigation.  Taking advantage of potential specific structures of $\bm Q_1$ and $\bm Q_2$, data augmentation strategies introduce one (or several) auxiliary variable $\bm u \in \mathbb R^d$ such that the joint distribution of the pair $(\bm u, \bm z)$ yields simple conditional distributions, thus sampling steps for a Gibbs sampler. One can recover the target distribution $\mathcal{N}(\bm \mu, \bm Q^{-1})$ via marginalization of the auxiliary variable $\bm u$, either exactly (as in exact data augmentation schemes, like GEDA) or in an asymptotic regime.

In GEDA, as described in Ref.~\cite{marnissi2019geda}, the underlying assumption is that the precision matrix $\bm Q$ can be split as follows:
\begin{equation}
\begin{split}
    & \bm Q = \bm Q_1 + \bm Q_2 \\
    & \bm Q_1 = \bm G_1^T \bm D_1 \bm G_1 \\
    & \bm Q_2 = \bm U_2^T \bm D_2 \bm U_2 \,,
\end{split}
\label{eq:precision_split}
\end{equation}
where $\bm G_1$ is arbitrary, $\bm D_1$ is diagonal and positive definite, $\bm U_2$ is unitary, and  $\bm D_2$ is diagonal. 
GEDA introduces two auxiliary variables, $\bm u_1$ and $\bm u_2$, such that the joint distribution is
\begin{equation}
\begin{split}
    p(\bm z, \bm u_1, \bm u_2) \propto & \exp\left(-\frac{1}{2}\left[(\bm z-\bm\mu)^T \bm Q(\bm z-\bm\mu) + (\bm u_1 - \bm z)^T \bm R(\bm u_1 - \bm z)\right] \right) \\
    &\times \exp\left( -\frac{1}{2}(\bm u_2-\bm G_1 \bm  u_1)^T \bm D_1(\bm u_2-\bm G_1 \bm  u_1)\right) \,,
\end{split}
\end{equation}
where $\bm R = \omega^{-1} \bm I_d - \bm Q_1$, and  $\omega$ is a positive hyper-parameter of the algorithm that must obey $1 < \omega < 1/ \lvert \bm Q_1 \rvert$, meaning that $1/\omega$ should be larger than the largest singular value of $\bm Q_1$. 

This joint distribution yields conditional Gaussian distributions with diagonal covariance matrices for both $\bm u_1$ and $\bm u_2$ that can be sampled efficiently by a Gibbs sampler with the following steps:
\begin{equation}
\begin{array}{ccl}
    1. & \bm u_2\sim\mathcal{N}(\bm \mu_{\bm u_2}, \bm Q^{-1}_{\bm u_2}) & \text{with } \bm \mu_{\bm u_2} = \bm G_1 \bm u_1 \text{ and }\bm Q_{\boldsymbol u_2} = \boldsymbol D_1, \\
    2.  & \bm u_1\sim\mathcal{N}(\bm \mu_{\bm u_1}, \bm Q^{-1}_{\bm u_1})  & \text{with } \bm \mu_{\bm u_1} = \bm z - \omega(\bm Q_1 \bm z - \bm G_1^T \bm D_1^{-1} \bm u_2) \text{ and } \bm Q_{\bm u_1} = \omega^{-1} \bm I_d, \\
    3.  & \bm u_1\sim\mathcal{N}(\bm \mu_{\bm u_1}, \bm Q^{-1}_{\bm u_1}) & \text{with }\bm \mu = \bm Q(\bm R \bm  u_1 + \bm  Q \bm \mu) \text{ and } \bm Q = \omega^{-1} \bm I_d + \bm Q_2 \;.
\end{array}
\end{equation}

As detailed in Sec.~\ref{sec:method}, in our setup the dimensionality is equivalent to the number of pixels $d=N^2$, the covariance matrix is $\bm Q = \bm Q_{\text{post}} = \bm Q_{\text{like}} + \bm Q_{\text{prior}}$, and $\bm G_1= \bm I$.

\section{Appendix: Second-order Lagrangian perturbation theory (2LPT)}
\label{apx:2lpt}
Second-order Lagrangian perturbation theory (2LPT) describes the cosmological dynamics of a cold, collisionless medium. It is based on a fluid description of the Vlasov--Poisson system (\textit{e.g.} \cite{Peebles2020}) and therefore ceases to be valid as soon as particle trajectories cross (so-called ``shell-crossing''). Although the density fields we consider herein are well in the post-shell-crossing regime, 2LPT still serves as a suitable forward model for validating our image reconstruction method.

The central quantity in 2LPT is the displacement $\bm \psi(\bm q, a) = \bm p(\bm q, a) - \bm q$, i.e.\ the vector pointing from each Lagrangian (``initial'') position $\bm q$ to the associated Eulerian position ${\bm p}({\bm q}, a)$ at scale-factor time $a$ on the characteristic curve originating at $\bm q$. LPT expands the displacement as a Taylor series w.r.t.\ $a$, with a purely space-dependent coefficient $\bm \psi^{(n)}(\bm q)$ at the $n$-th order. For 2LPT, this yields
\begin{equation}
    \bm \psi(\bm q, a) = a \bm \psi^{(1)}(\bm q) + a^2 \bm \psi^{(2)}(\bm q) \;,
\end{equation}
where 
\begin{subequations}
\begin{align}
    \bm \psi^{(1)} &= - \bm \nabla_{\bm q} \phi^0 \;, \\
    \bm \psi^{(2)} &= -\frac{3}{7} \bm \nabla_{\bm q} \, \Delta_{\bm q}^{-1} \left[(\partial_{q_1, q_1} \phi^0) \, (\partial_{q_2, q_2} \phi^0) - (\partial_{q_1, q_2} \phi^0)^2 \right] \;.
\end{align}
\end{subequations}
Here, $\phi^0$ is the initial gravitational potential, and the scale factor $a$ simply acts as a time variable and has no physical meaning in 2D.  Note that we consider Einstein--de Sitter cosmology ($\Omega_m = 1$); for extensions to the $\Lambda$CDM case and higher-order LPT, see \textit{e.g.} Ref.~\cite{Rampf2022} and references therein.

Given the 2LPT displacement $\bm \psi$, the density contrast $\delta$ is given by
\begin{equation}
    1 + \delta(\bm p(\bm q, a)) = \int \delta_\text{D}(\bm p(\bm q, a) - \bm q - \bm \psi(\bm q, a)) \ \mathrm{d}^2q \;,
\end{equation}
where $\delta_\text{D}$ is the Dirac $\delta$-distribution. In practice, we use cloud-in-cell interpolation \cite[\textit{e.g.}][]{HockneyEastwood} to compute the density contrast on a Cartesian grid. Finally, let us remark that since we view the discrete density contrast as an $N \times N$-dimensional image in the main part of this work, we use the bold symbol $\bm \delta$ in that context.

\end{appendix}

\clearpage


\end{document}